\documentclass[preprint,twocolumn,5p,authoryear,colorlinks=true]{elsarticle}

\usepackage{lineno,hyperref}
\modulolinenumbers[5]

\journal{Icarus}

\usepackage[labelfont=bf,justification=justified,singlelinecheck=false]{caption}
\usepackage{numcompress}

\begin{document}

\begin{frontmatter}

\title{Radar scattering of linear dunes and mega-yardangs: Application to Titan}

%% Group authors per affiliation:
\author[LAB]{Philippe Paillou\corref{correspondingauthor}}
\ead{philippe.paillou@obs.u-bordeaux1.fr}
\author[URCA]{Beno\^{i}t Seignovert}
\author[BYU]{Jani Radebaugh}
\author[JPL]{Stephen Wall}

\address[LAB]{Universit\'{e} de Bordeaux, UMR 5804-LAB, 33270 Floirac, France}
\address[URCA]{Universit\'{e} de Reims, UMR 7331-GSMA, 51687 Reims, France}
\address[BYU]{Brigham Young University, Department of Geological Sciences, Provo, UT 84602, USA}
\address[JPL]{Jet Propulsion Laboratory, California Institute of Technology, Pasadena, CA 91109, USA}

\cortext[correspondingauthor]{Corresponding author}

\begin{abstract}
The Ku-band (13.8 GHz - 2.2 cm) RADAR instrument onboard the Cassini-Huygens spacecraft
has revealed the richness of the surface of Titan, as numerous seas, lakes, rivers,
cryo-volcanic flows and vast dune fields have been discovered. Linear dunes are a major
geomorphological feature present on Titan, covering up to 17\% of its surface,
mainly in equatorial regions. However, the resolution of the RADAR instrument is not good
enough to allow a detailed study of the morphology of these features. In addition, other
linear wind-related landforms, such as mega-yardangs (linear wind-abraded ridges formed in
cohesive rocks), are likely to present a comparable radar signature that could be confused
with the one of dunes. We conducted a comparative study of the radar radiometry of both
linear dunes and mega-yardangs, based on representative terrestrial analogues: the linear
dunes located in the Great Sand Sea in western Egypt and in the Namib Desert in Namibia,
and the mega-yardangs observed in the Lut Desert in eastern Iran and in the Borkou Desert
in northern Chad. We analysed the radar scattering of both terrestrial linear dunes and
mega-yardangs, using high-resolution radar images acquired by the X-band (9.6 GHz - 3.1
cm) sensor of the TerraSAR-X satellite. Variations seen in the radar response of dunes are
the result of a contrast between the dune and interdune scattering, while for
mega-yardangs these variations are the result of a contrast between ridges and erosion
valleys. We tested a simple surface scattering model, with parameters derived from the
local topography and surface roughness estimates, to accurately reproduce the radar signal
variations for both landforms. It appears that we can discriminate between two types of
dunes - bare interdunes as in Egypt and sand-covered interdunes as in Namibia, and between
two types of mega-yardangs - young yardangs as in Iran and older ones as in Chad. We
applied our understanding of the radar scattering to the analysis of Cassini RADAR T8
acquisitions over the Belet Sand Sea on Titan, and show that the linear dunes encountered
there are likely to be of both Egyptian and Namibian type. We also show that the
radar-bright linear features observed in Cassini RADAR T64 and T83 acquisitions are very
likely to be mega-yardangs, possible remnants of ancient lake basins at mid-latitude,
formed when Titan’s climate was different.
\end{abstract}


\end{frontmatter}

%\linenumbers

\section{Introduction}
Cassini-Huygens mission has been in orbit around Saturn since June
2004. The Ku-band (13.8 GHz - 2.2 cm) RADAR instrument onboard the
Cassini spacecraft is a combined
radiometer/altimeter/scatterometer/imaging radar that has revealed a
various and rich surface of Titan through its optically-opaque
atmosphere \citep{Elachi2004}. RADAR has allowed the discovery of
numerous seas, lakes, rivers, cryo-volcanic structures and vast dune
fields
\citep{Elachi2005,Lopes2007,Lorenz2006,Radebaugh2007,Stofan2007}. Dunes
are in particular a major landform on the surface of Titan, since
large dune fields cover more than 10 million km$^2$ in equatorial
regions. They are typically 1-2 km wide, with 1-4 km spacing, up to
150 m-high, and can reach more than 100 km in length, being mainly
east-west oriented and aligned parallel with time-averaged equatorial
winds \citep{Lorenz2006,Radebaugh2008,Lorenz2009,LeGall2011}. Dunes on
Titan are the linear type as observed on Earth, the latter ones being
used as analogues to infer Titan’s dunes morphology
\citep{Neish2010,Radebaugh2010,Paillou2014}.

The RADAR instrument shows Titan’s dune fields as dark linear features
separated by brighter linear features. Different qualitative
interpretations have been proposed for this radar signature: (1) dark
lines are sand covered interdunes with brighter features caused by
specular reflection over dunes’ crests, as for the linear dunes
observed in the Namib Desert \citep{Neish2010}, or (2) dark lines are
the smooth dunes with brighter linear features caused by rougher
interdunes, where bedrock is exposed, as observed in the Great Sand
Sea in Egypt \citep{Paillou2014}.  Besides linear dunes, other natural
wind-related structures on Earth show a comparable morphology:
mega-yardangs are wind-abraded landforms which develop in Earth’s
drylands (Sahara, Middle-East, Central Asia), where winds tend to be
unimodal in direction \citep{Goudie2007}.  Mega-yardangs are composed
of alternating linear ridges and valleys created by wind erosion and
sediment deflation, and are often associated with soft deposits of
desiccated lake beds. As for linear dunes, radar images of
mega-yardang structures show alternating dark linear features (the
erosion valleys) and brighter linear features (the ridges): confusing
radar images of linear dunes and mega-yardangs is then quite possible,
especially at the 300 m resolution of the RADAR instrument.

Comparative planetology is a powerful approach to help understand the
geology of remote planetary surfaces. We conducted a comparative study
of the radar scattering of both linear dunes and mega-yardangs, based
on representative terrestrial analogues. We considered the linear
dunes located in the Great Sand Sea in western Egypt and in the Namib
Desert in Namibia, and the mega-yardangs located in the Lut Desert in
eastern Iran and in the Borkou Desert in northern Chad. We analysed
and modelled the radar signatures of both linear structures using
high-resolution (18 m) radar images acquired by the X-band (9.6 GHz -
3.1 cm) radar of the TerraSAR-X satellite. We used a simple surface
scattering model, whose parameters were derived from the local
topography (ASTER Global Digital Elevation Map - GDEM, and Shuttle
Radar Topography Mission - SRTM data) and from surface roughness
estimates (Bayesian inversion), which accurately reproduces the radar
signature for both landforms. We were able to discriminate between two
types of dunes: those with bare interdunes in Egypt vs. those with
sand-covered interdunes in Namibia; and between two types of
mega-yardangs: young ones in Iran vs. older ones in Chad.

We applied our understanding of the radar scattering to the analysis
of radar images obtained during the Cassini T8 flyby over the Belet
Sand Sea on Titan: we show that the linear dunes there are likely to
be of both Egyptian and Namibian types, contradicting previous studies
which proposed single-type dune scenarios. We also show that the
bright linear structures observed in radar acquisitions during Cassini
T64 and T83 flybys are very likely to be mega-yardangs, possible
remnants of lake beds at mid-latitude (around 40$^\circ$N).

\section{Linear dunes and mega-yardangs on Earth}
Linear dunes on Earth are mainly located in arid equatorial regions,
and constitute more than half of terrestrial dunes \citep{Rubin2009}.
They are formed in a context of moderate sand supply, with bimodal
wind regimes associated with seasonal changes, producing
quasi-symmetrical and linear dunes in a direction parallel to the one
of average annual wind \citep{Bristow2000}.  Linear dunes can reach
hundreds of kilometres in length, and more than a hundred metres in
height, with an interdune separation of the order of a couple of
kilometres \citep{Besler2008}.  Such results of sedimentary transport
and deposition processes have been observed on all bodies of the Solar
System that have an atmosphere: Venus, Earth, Mars and Titan
\citep{Zimbelman2013}.

Also belonging to the “wind-related linear features” family, are
yardangs (or mega-yardangs at a regional scale), made of parallel
ridges separated by narrow valleys. The latter are formed by the wind
abrasion of cohesive rocks \citep{Goudie2007}.  Mega-yardangs are also
mainly located in hyperarid regions on Earth, and form under the
condition of a strong unimodal wind, that transports sand and gravel
eroding soft sediments on a time scale of a million years
\citep{Gabriel1938,Ehsani2008}.  A favourable condition for yardangs
formation is the presence of soft deposits over a harder bedrock,
typically observed in areas of ancient lake basins. Mega-yardangs were
observed on Mars \citep{DeSilva2010} and might also be present on
Venus \citep{Greeley1999}.

We selected four terrestrial sites to study radar scattering of linear
dunes and mega-yardangs. For linear dunes, we considered the Egyptian
side of the Great Sand Sea, a large dune field covering 300 x 700 km
in eastern Libya and western Egypt \citep{Besler2008}.  The typical
length, width and height of dunes in the Great Sand Sea are in the
range of values proposed for Titan, the interdune being most of the
time uncovered and exposing the underlying bedrock. Egyptian dunes are
mainly composed of pure silicate, with a low dielectric constant close
to the values reported for Titan’s surface materials
\citep{Paillou2008}.  They have been used as terrestrial analogues to
develop a scattering model that was applied to the study of the radar
response of Titan’s dunes \citep{Paillou2014}. The Namib
Desert in Namibia \citep{Lancaster1989,Bristow2000} was also
considered for its linear dune fields. It is an arid area which
borders the southwestern African coast over hundreds of kilometres,
with mostly linear sand dunes presenting sand-covered interdunes, and
also comparable in size and morphology to those on Titan. These dunes
have been used as a training model for radarclinometry (reconstruction
of the topography from the radar signature), with results applied to
the study of Titan’s dunes \citep{Neish2010}.

As regards mega-yardangs, we considered the young structures of the
Lut Desert in eastern Iran, firstly described by \cite{Gabriel1938}
and later studied by \cite{Ehsani2008}.  The region corresponds to a
Pleistocene basin with fill deposits (silty clay, gypsiferous sands)
and is about 150 km long by 50 km wide, with ridges reaching heights
of 100 m \citep{Goudie2007}.  We also considered the older
mega-yardangs of the Borkou Desert in northern Chad, formed by erosion
of sandstone of Paleozoic age \citep{Mainguet1968}.  This is a region
located between the Tibesti and Ennedi mountains, in a corridor of
very strong unimodal wind oriented to the northeast. Eroded valleys
between sandstone ridges are mostly filled with aeolian sand and
present an average depth around 50 m \citep{McHone1996}.

We acquired radar scenes of the four study sites using the X-band
radar of the TerraSAR-X satellite, launched in 2007 \citep{Pitz2010}.
The TerraSAR-X sensor presents two interesting characteristics for our
study: (1) X-band is the shortest wavelength (3.1 cm) among available
orbital radars for Earth observation, close enough to the Cassini
RADAR (2.2 cm) and thus sensitive to the same surface roughness range
and (2) its high-resolution (18 m) allows the detailed study of the
radar signatures of landforms. We used the TerraSAR-X ScanSAR mode, in
HH polarisation, to cover 100 x 150 km areas at an average resolution
of 18 m (multi-look images resampled to 8.25 m), with a varying
incidence angle from the near to the far range of the
image. Acquisition parameters of the four TerraSAR-X scenes are
summarized in Table~\ref{tab:1}.

\begin{table*}
  \centering
  \caption{\label{tab:1}The four ScanSAR scenes acquired using the TerraSAR-X radar,
    orbit inclination is 97$^\circ$.}
  \small
  \begin{tabular}{l c c c c l}
\hline
Site - Country       & Scene centre Lat./Lon. & Incidence angle range & Acquisition date & Scene ID & Orbit look \\
\hline
Great Sand Sea Egypt & $24^\circ 30^\prime$N $26^\circ 10^\prime$E & $19.5-30.0^\circ$ & 2013-06-06 & 371348705\_2 & ascending right \\
Namib Desert Namibia & $24^\circ 30^\prime$S $15^\circ 10^\prime$E & $24.5-34.5^\circ$ & 2013-12-27 & 372338466\_1 & ascending right\\
Lut Desert Iran & $30^\circ 15^\prime$N $58^\circ 20^\prime$E & $29.0-38.5^\circ$ & 2013-06-08 & 371348705\_4 & ascending right\\ 
Borkou Desert Chad & $19^\circ 00^\prime$N $19^\circ 05^\prime$E & $24.5-35.0^\circ$ & 2013-12-08 & 372239342\_1 & ascending right\\ 
\hline
\end{tabular}
\end{table*}

\begin{figure}
  \centering
  \includegraphics{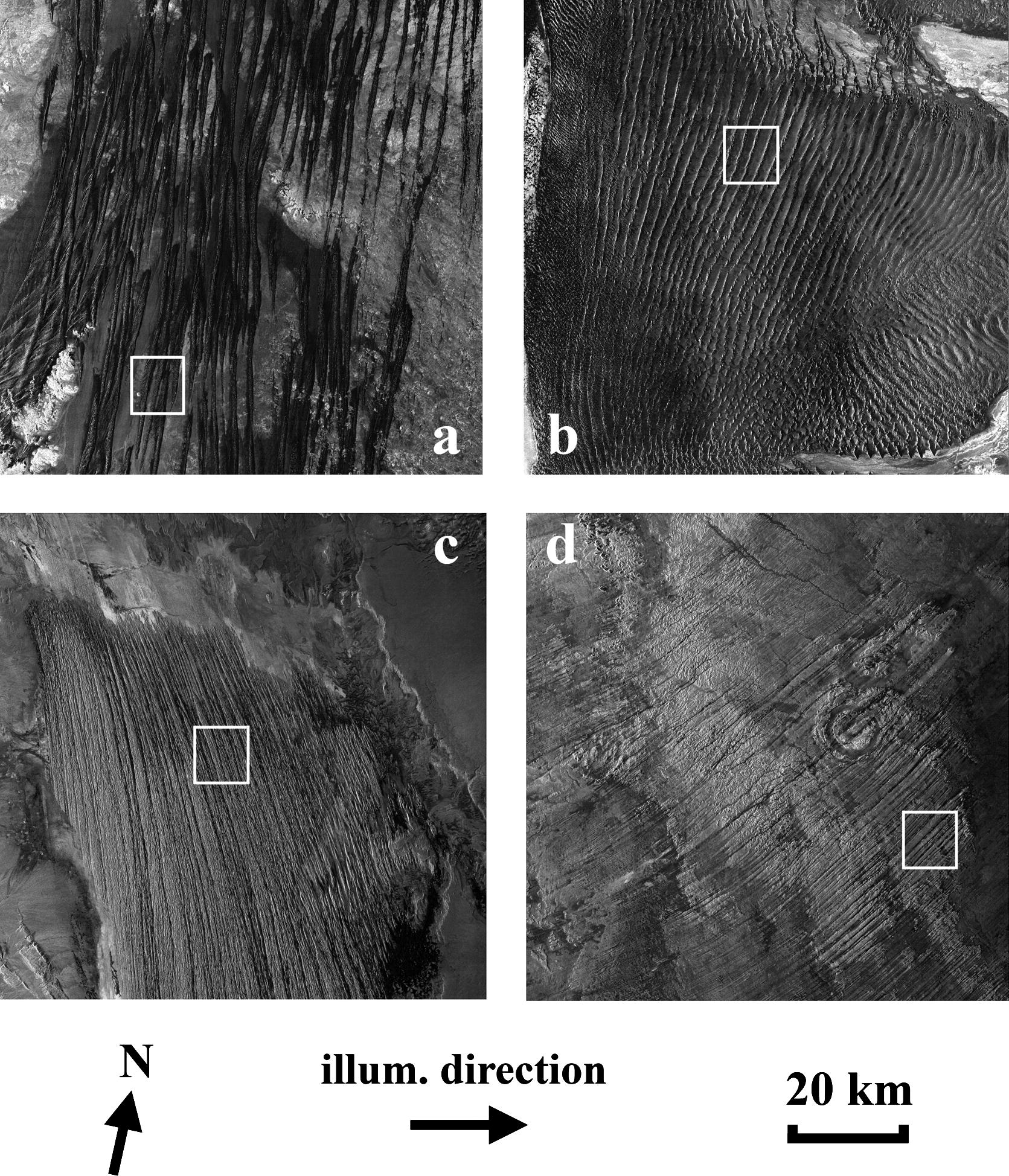}
  \caption{\label{fig:1}Four TerraSAR-X ScanSAR scenes of linear dunes
of (a) the Great Sand Sea in Egypt, (b) the Namib Desert in Namibia;
mega-yardangs of (c) the Lut Desert in Iran and (d) the Borkou Desert
in Chad. See Table.~\ref{tab:1} for geographical coordinates, white
squares show locations of full resolution images presented in
Fig.~\ref{fig:2}.}
\end{figure}

\begin{figure}
  \centering
  \includegraphics{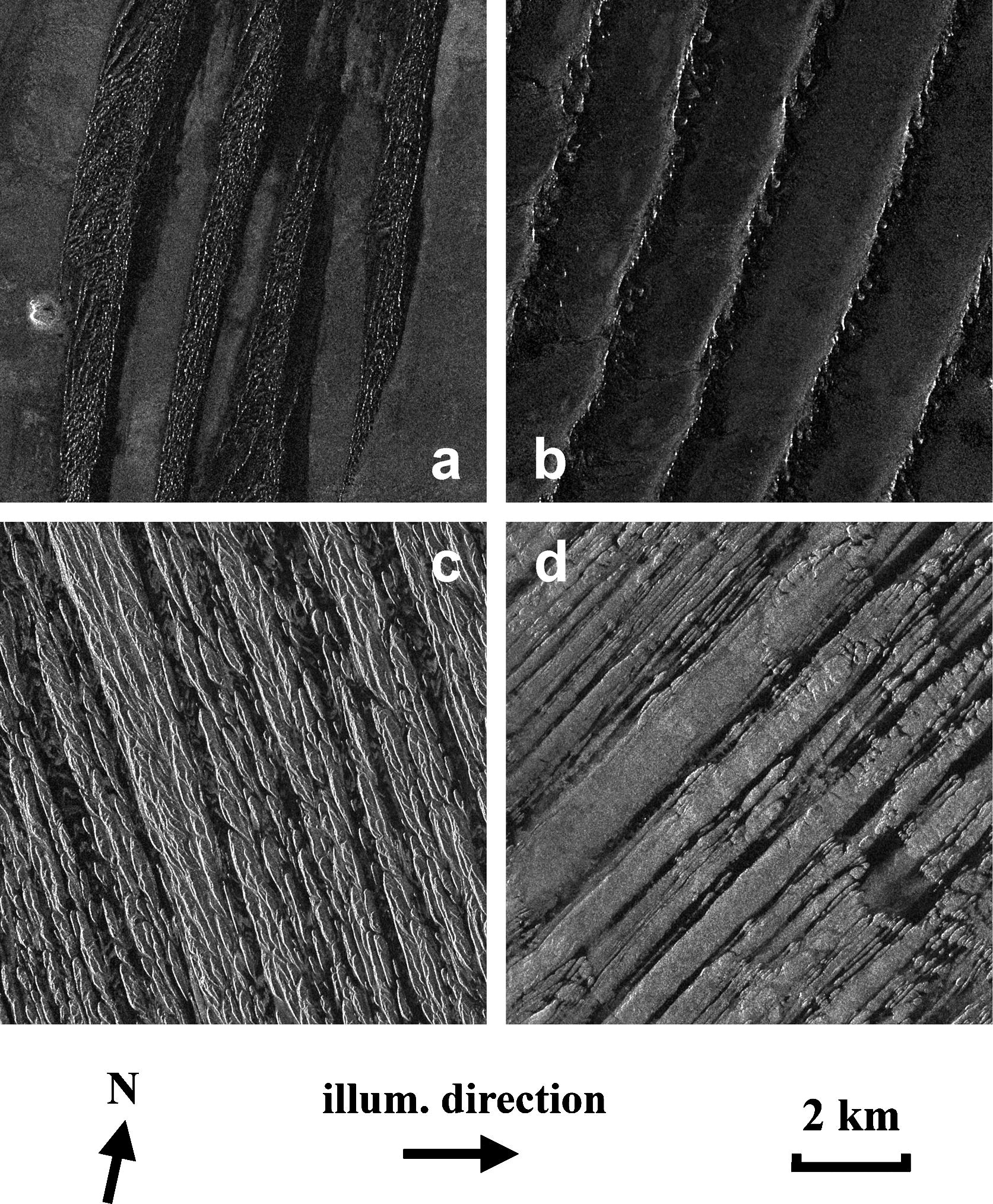}
  \caption{\label{fig:2}Full resolution (8.25 m/pixel) extracts of the
TerraSAR-X scenes: linear dunes of (a) the Great Sand Sea in Egypt
(incidence angle 23.0$^\circ$), (b) the Namib Desert in Namibia
(incidence angle 29.2$^\circ$); mega-yardangs of (c) the Lut Desert in
Iran (incidence angle 33.6$^\circ$) and (d) the Borkou Desert in Chad
(incidence angle 33.5$^\circ$).}
\end{figure}

The TerraSAR-X radar scenes are shown in Fig.~\ref{fig:1}, with
relevant full resolution extracts presented in Fig.~\ref{fig:2}.  Both
linear dunes of the Great Sand Sea and Namib Desert are roughly
north–south oriented and are comparable in size, but they show a
different radar signature mainly due to the interdune properties: the
bare interdune in Egypt (Fig.~\ref{fig:1}a) appears brighter than the
sand-covered one in Namibia (Fig.~\ref{fig:1}b), because of the
rougher surface associated with the exposed bedrock, which causes a
more diffuse scattering of the incident radar wave. Full resolution
extracts in Fig.~\ref{fig:2}a for Egyptian dunes and Fig.~\ref{fig:2}b
for Namibian dunes respectively show some brighter radar returns on
the side of the dunes facing the radar illumination, due to the
combination of a lower local incidence angle \citep{Blom1987} and
finer dune structures, some of them acting as natural corner
reflectors \citep{Paillou2014}.  TerraSAR-X's high resolution also
allows us to see the darker ``back side'' of the linear dunes,
corresponding to a low radar return because of a higher local
incidence angle.

Mega-yardangs of the Lut Desert and of the Borkou Desert do not
present the same geographical orientation (Fig.~\ref{fig:2}c and d),
but both appear generally brighter than linear dunes: most of the
radar scattering is due to the wide ridges of the eroded layer, which
is rougher than sand deposits at X-band. Narrow linear valleys
separate the ridges and appear darker because they are filled with
smoother aeolian deposits (some valleys even contain very small
dunes). The young mega-yardangs of the Lut Desert show complex ridges
structures alternating with poorly defined valleys (see
Fig.~\ref{fig:2}c), while the older mega-yardangs of the Borkou Desert
exhibit a better contrast between wider and flatter ridges and deeper
erosion valleys (see Fig.~\ref{fig:2}d).

\section{Description and modelling of the radar signature of dunes and yardangs}
We studied the variations of the radar signal across selected dune and
yardang structures, at TerraSAR-X full-resolution, in order to
qualitatively understand the scattering mechanisms involved, and then
to reproduce the observed variations with the help of a simple surface
scattering model.

\subsection{Radar signatures of linear dunes and mega-yardangs}
\begin{figure*}
  \centering
  \includegraphics{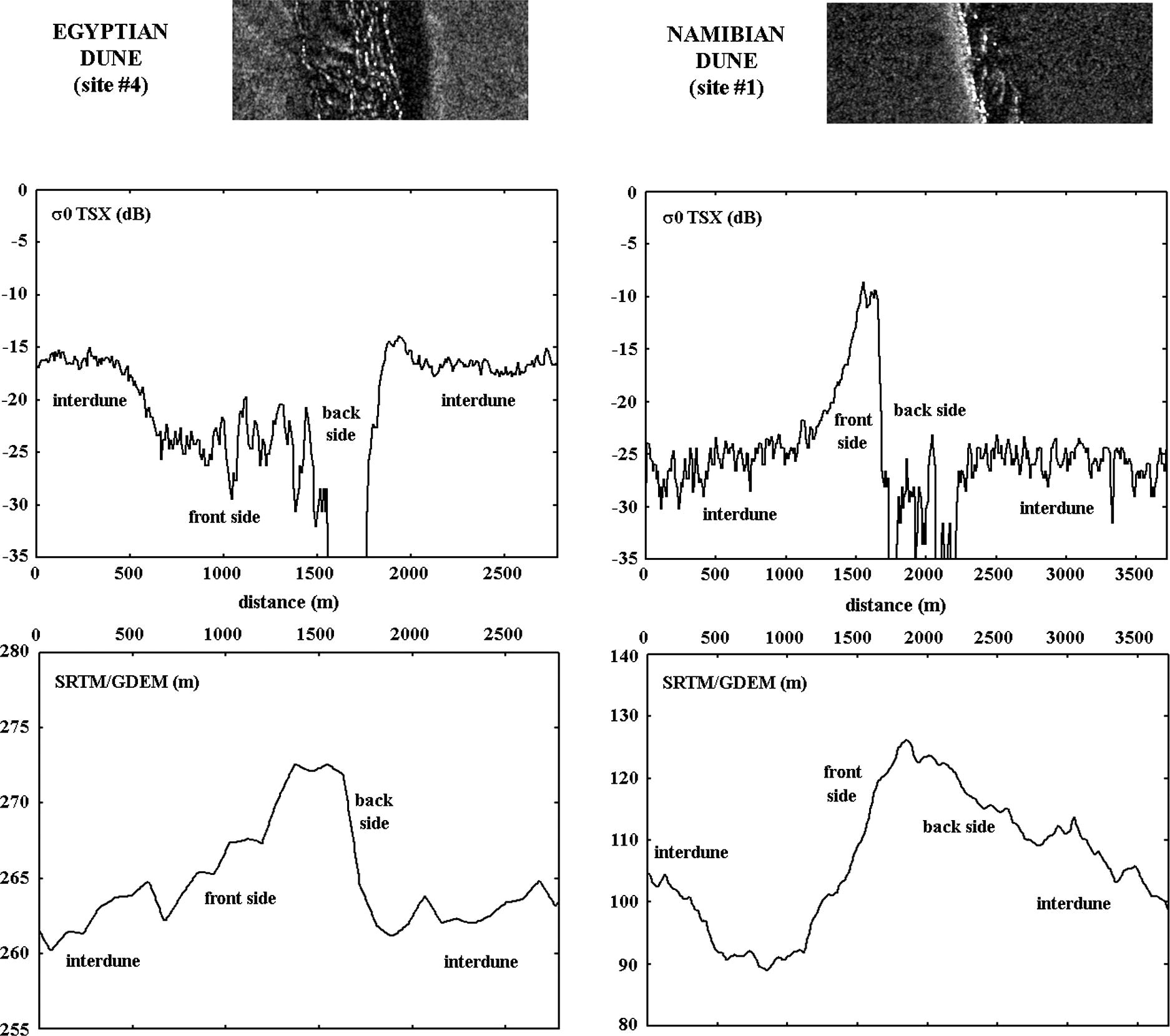}
  \caption{\label{fig:3}TerraSAR-X radar scattering and topography (SRTM
with voids filled using GDEM) profiles across a linear dune of the
Great Sand Sea in Egypt (left) and of the Namib Desert in Namibia
(right). The analysed full resolution TerraSAR-X extracts are shown on
top of the figure (site \#4 for the Egyptian dune, at incidence angle
of 23$^\circ$, and site \#1 for the Namibian dune, at incidence angle
of 29.2$^\circ$), north is up and radar illumination is from the
left. Vertical axis of plots is backscattered power (in dB) or
altitude (in m) while horizontal axis is the distance across the dunes
(in m). Location of interdune, front side and back side of the dunes
are indicated on each plot.}
\end{figure*}

Eight representative locations were selected in the Great Sand Sea and
six in the Namib Desert, in order to derive a ``typical'' radar
scattering profile for Egyptian and Namibian dunes. We applied a
standard calibration procedure to full resolution TerraSAR-X data to
compute the backscattered radar power at each pixel
\citep{Schwerdt2010}.  We averaged pixels values in the direction
parallel to the main orientation of the structures, in order to get a
better representation of typical variations of the radar scattering
across the studied structures: averaging was performed using a median
estimator, so that extreme high returns due to natural corner
reflectors and extreme low returns at the noise floor level (around
-28 dB for TerraSAR-X) did not bias the average
value. Fig.~\ref{fig:3} shows two typical radar signatures across
Egyptian and Namibian linear dunes. Egyptian dunes present three main
levels of radar scattering (Fig.~\ref{fig:3}, left): a radar-bright
interdune (around -15 dB) due to an exposed rough bedrock
\citep{Besler2008}, a medium scattering level for the dune’s side
facing the radar illumination (``front side'', around -23 dB), and a
low scattering level for the dune’s side opposite to the radar
illumination (``back side'', -30 dB and lower). Namibian dunes present
a different, rather bimodal, radar signature, with a less radar-bright
interdune due to a smooth sand cover
\citep{Bristow2000}. Fig.~\ref{fig:3} (right) shows a typical radar
profile across a Namibian dune, where two main scattering levels can
be observed: a strong radar return on the side of the dune facing the
radar, can reach -10 dB (due to a lower incidence angle combined with
smaller-scale structures superposed to the main dune’s shape
\citep{Lancaster1989}.  A lower scattering level, in the -25 to -30 dB
range, corresponds to returns from both interdune and back side of the
dune (the latter likely mixing with the interdune return when
decreasing the radar image resolution). This shows that, while both
dune systems in Egypt and Namibia are of the same linear type, their
radar signature can be quite different due to different interdune
properties. This appears quite obvious here from high-resolution radar
images, but at the lower resolution (and higher noise level) of the
Cassini RADAR instrument, observing these structures can become more
confusing.

\begin{figure*} \centering
  \includegraphics{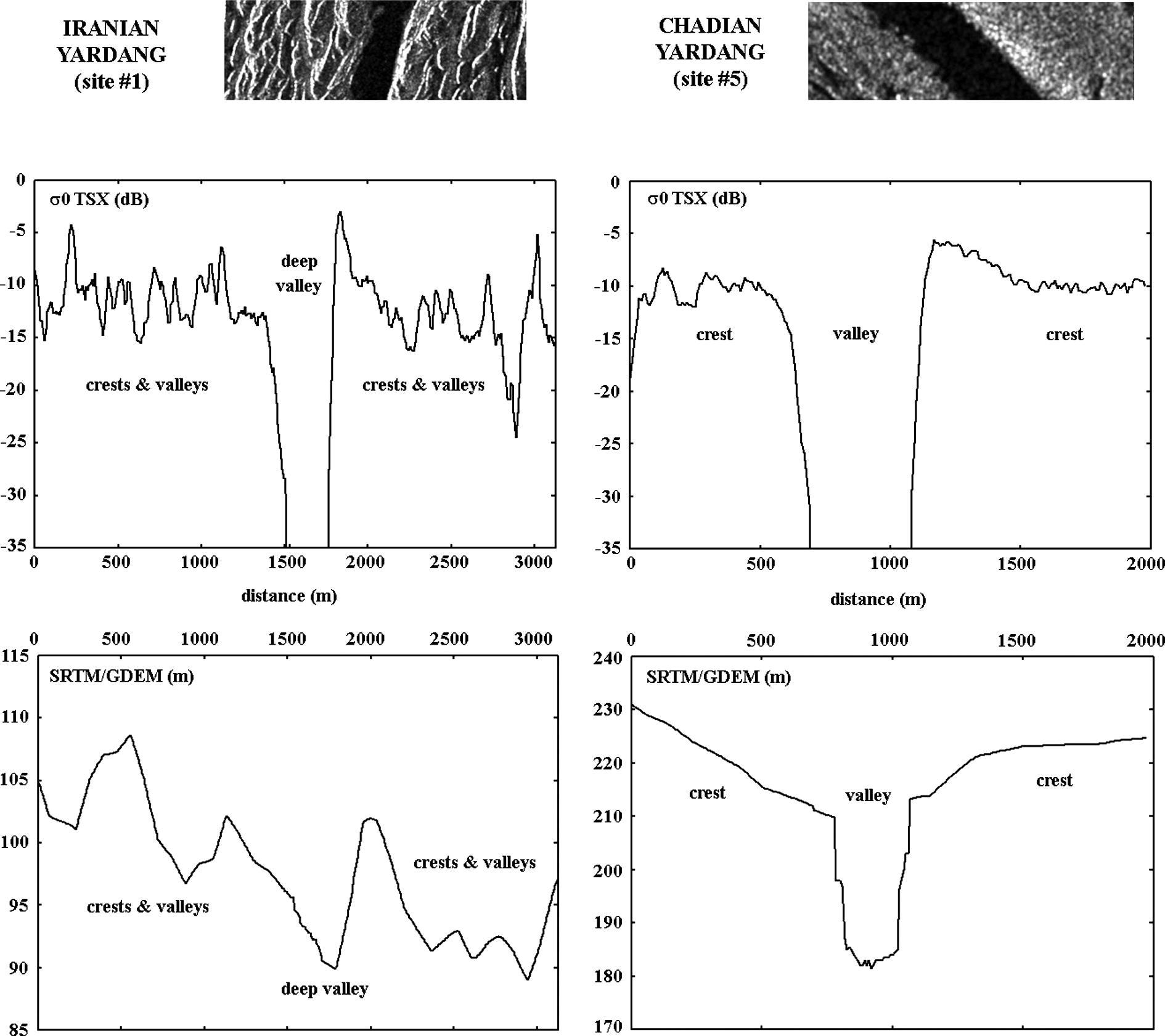}
  \caption{\label{fig:4}TerraSAR-X radar scattering and topography
(SRTM with voids filled using GDEM) profiles across mega-yardangs of
the Lut Desert in Iran (left) and of the Borkou Desert in Chad
(right). The analysed full resolution TerraSAR-X extracts are shown on
top of the figure (site \#1 for Iranian yardangs, at incidence angle of
33.6$^\circ$, and site \#5 for Chadian yardangs, at incidence angle of
33.5$^\circ$), north is up and radar illumination is from the
left. Vertical axis of plots is backscattered power (in dB) or
altitude (in m) while horizontal axis is the distance across the
yardangs (in m). Location of valleys and crests are indicated on each
plot.}
\end{figure*}

Regarding mega-yardangs, we studied five representative
locations in the Lut Desert and also five in the Borkou Desert, to
derive typical radar profiles for young and older mega-yardangs. As
before, we averaged the pixel values in the direction parallel to the
main orientation of the structures using a median
estimator. Fig.~\ref{fig:4} presents two typical radar signatures
across Iranian and Chadian yardangs. The young mega-yardangs in Iran
(Fig.~\ref{fig:4}, left) show a high spatial frequency alternation of
the high scattering level from the rough crests of the eroded layer
($>$-10 dB) and the medium-scattering level from narrow and shallow
valleys (around -15 dB).  This alternation is cut in places by regions
of low radar return ($<$-25 dB), due to deeper eroded valleys filled
with smooth aeolian deposits \citep{Gabriel1938}. The older
mega-yardangs in Chad present a different, more bimodal radar
signature (Fig.~\ref{fig:4}, right), with alternation of radar-bright
level from wider and flatter crests (around -10 dB, less radar-bright
than the Iranian ones likely because a longer period of erosion
produced a smoother surface), and low-scattering level due to larger
erosion valleys ($<$-25 dB), which are likely to be filled with
aeolian sand deposits. As for linear dunes, the two mega-yardangs
structures, of different age and created in different types of rocks,
exhibit differences in their radar signature.  However, it is apparent
that mega-yardangs in general are more radar-bright than linear dunes,
since we observe about 10 dB difference in the average backscattered
power. This is likely due to the strong radar return of the rougher
and wider yardangs’ crests, compared with a lower overall signal
returned by the smoother sandy surfaces of dunes.

\subsection{A simple model for the radar signature of dunes and yardangs}
After presenting a qualitative description of radar scattering
profiles across linear dunes and mega-yardangs, we tried to reproduce
the variations of the observed radar backscattered power using simple
surface scattering models, whose parameters were estimated from
various sources. We considered two surface scattering models, widely
used by the radar remote sensing community: the Geometric Optics Model
– GOM – suited to rough surfaces (surface roughness being defined with
respect to the radar wavelength, most natural surfaces are rough at
X-band) \citep{Fung1981}, and the Integral Equation Model – IEM –
valid for medium-rough to smooth surfaces \citep{Fung1992}.  The
latter model was previously used to reproduce the radar scattering of
linear dunes of the Great Sand Sea in Egypt, using SIR-C/X-SAR scenes
\citep{Paillou2014}.

Besides the radar wavelength and polarisation, input parameters for
these models are local incidence angle $\theta$ (computed from the
radar look angle and local slope at the observed point), surface
roughness described by rms-height $\sigma$ and correlation length $L$
(assuming a Gaussian autocorrelation function), and the dielectric
constant $\varepsilon$ of the material constituting the surface. The
dielectric constant of dry sediments and sedimentary rocks in desert
regions does not vary much and is the 3–6 range for its real part,
with an imaginary part close to zero \citep{Ulaby1990}. We fixed the
dielectric constant to that of silicate ($\varepsilon$ = 3.5) for all
surfaces. Note that small changes in this parameters are not important
when computing radar backscattered power, as compared to the effects
of variations in the incidence angle and surface roughness.  The
topography of the various selected areas was obtained using SRTM and
GDEM data, GDEM being mainly used to fill in holes in the SRTM
coverage, especially in dune areas. Taking into account the look angle
and orbit inclination of the TerraSAR-X radar, we computed slope maps
in the radar range direction, from which we derived the local
incidence angle $\theta$ at each pixel. The resolution of digital
elevation models provided by SRTM (90 m) is coarser than the
resolution of the TerraSAR-X images, so it was not possible to
reproduce scattering effects due to small-scale features, in
particular natural corner reflectors. As regards surface roughness
parameters, only the ones for the surface of Egyptian dunes were
estimated by previous studies \citep{Paillou2014}.  We then considered
a Bayesian inversion approach to estimate the surface roughness of
dunes and interdunes for both Egyptian and Namibian cases, and of
crests and valleys for both Iranian and Chadian yardangs.  We computed
the probability $P$ of similarity between the actual radar
backscattered power in TerraSAR-X images $\sigma_{TX}^0$ and the
computed radar backscatter power $\sigma_{th}^0$, for roughness
parameters $\sigma$ varying between 0.02 and 0.5 cm and $L$ varying
between 0.2 and 3.5 cm, assuming a tolerance criteria $\tau$= 0.03:

\begin{equation}
P(\sigma,L) = \frac{1}{\tau}
\exp{\left(-\left(\frac{\sigma_{TX} ^0 -
\sigma_{th}^0\left(\sigma,L\right)}{\tau^2}\right)^2\right)}
\end{equation}

\begin{figure} \centering
  \includegraphics{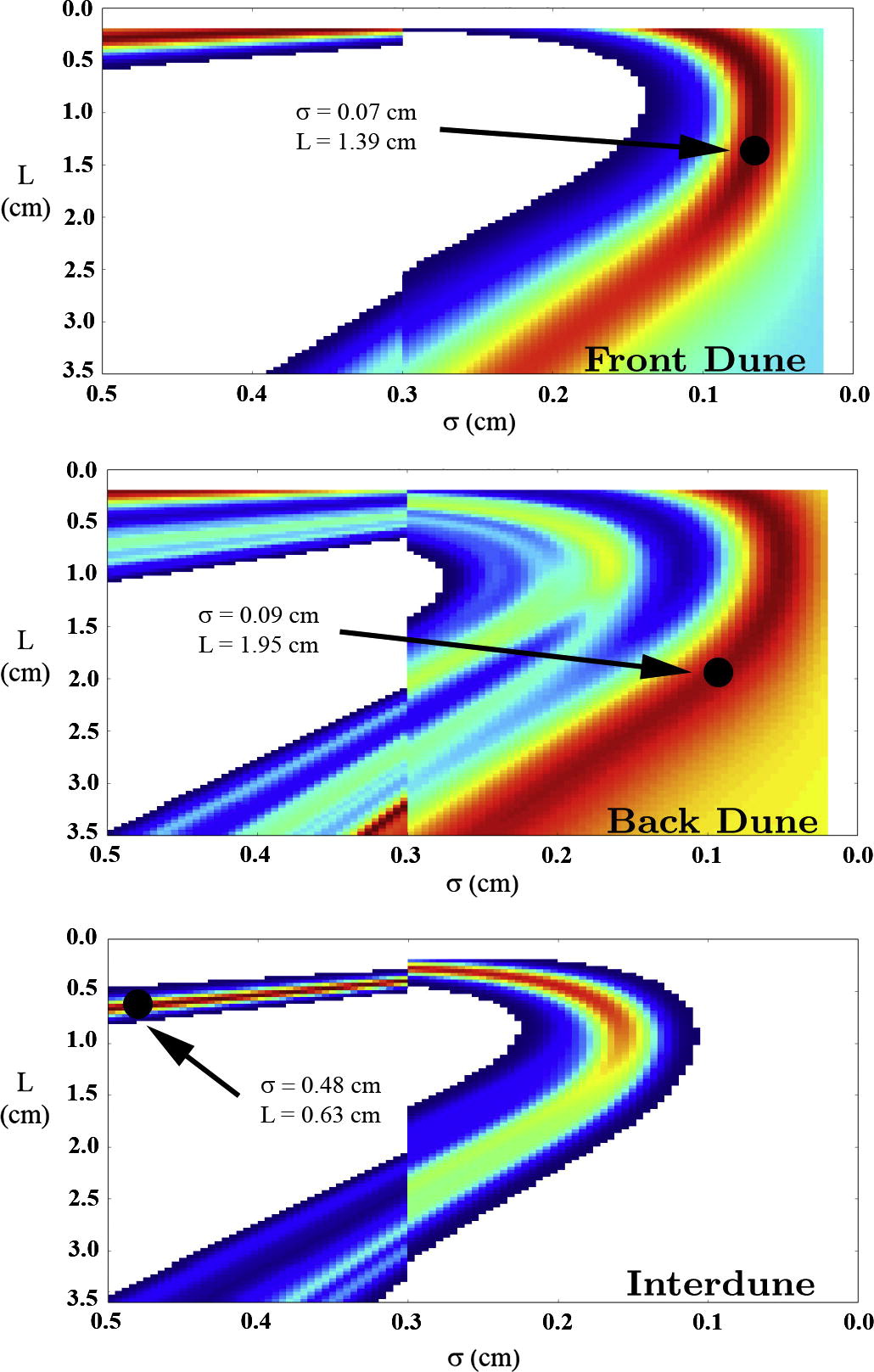}
  \caption{\label{fig:5} Bayesian inversion for the surface roughness
of an Egyptian dune. The probability $P$ increases from blue to red,
and the black dot is the location of the maximum probability. Top:
side of the dune facing the radar illumination (maximum at $\sigma$=
0.07 cm and $L$ = 1.39 cm). Middle: side of the dune opposite to the
radar illumination (maximum at $\sigma$= 0.09 cm and $L$ = 1.95
cm). Bottom: rough interdune (maximum at $\sigma$= 0.48 cm and $L$ =
0.63 cm). The discontinuity in probability at $\sigma$= 0.3 cm is due
to the transition from GOM to IEM model. (For interpretation of the
references to colour in this figure legend, the reader is referred to
the web version of this article.)}
\end{figure}

The GOM model was used when $\sigma >$ 0.3 cm and the IEM model was
used when $\sigma <$ 0.3 cm. After exploring the roughness parameter
space, we kept the ($\sigma$, $L$) couple which corresponds to a
maximum probability of similarity between the actual TerraSAR-X and
the computed backscattered power. Fig.~\ref{fig:5} shows an example of
this Bayesian approach for dunes and interdunes of the Great Sand Sea
in Egypt.

\begin{figure*} \centering
  \includegraphics{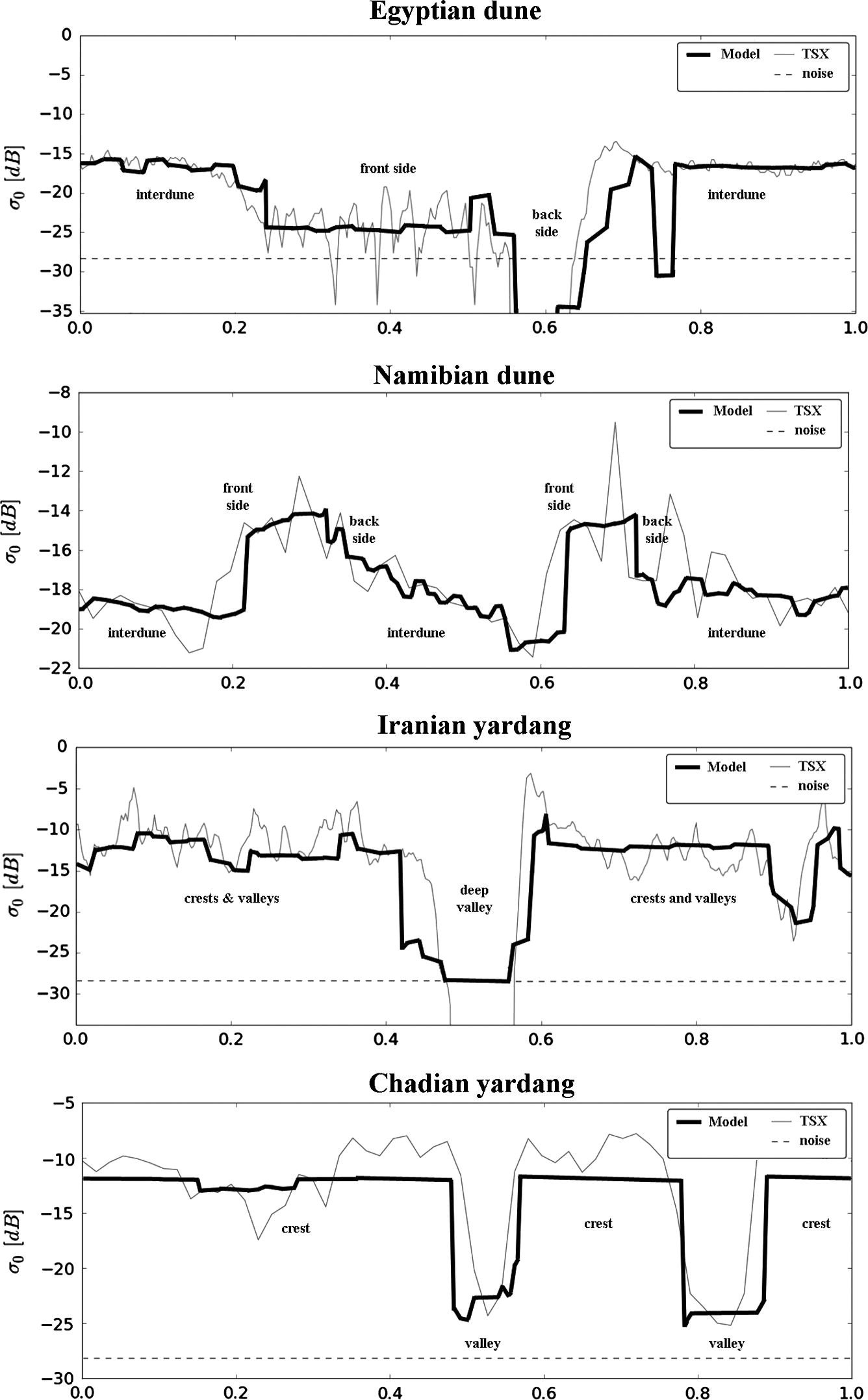}
  \caption{\label{fig:6}Comparison between radar scattering profiles
computed using our surface scattering model (bold black line) and
TerraSAR-X actual data (grey line). From top to bottom: Egyptian dune,
Namibian dune, Iranian yardang, and Chadian yardang, corresponding to
the sites shown in Fig.~\ref{fig:3} and Fig.~\ref{fig:4}. Vertical
axis is the backscattered power (in dB) and \emph{x}-axis is a
normalised distance across the studied structure. The dash-line
represents the TerraSAR-X noise level (around -28 dB).}
\end{figure*}

Once incidence angle and surface roughness parameters were estimated,
we computed ``theoretical'' radar scattering profiles using GOM
and IEM models, and compared them to observations for both linear
dunes and mega-yardangs. Fig.~\ref{fig:6} shows examples of comparison
between computed and observed radar profiles across linear dunes in
Egypt and Namibia, and mega-yardangs in Iran and Chad. One can see
that the computed radar profiles and the actual TerraSAR-X ones are
quite similar on the average. The high-frequency variations in
TerraSAR-X profiles could not be reproduced, due to the coarse
resolution of the SRTM topography. Nevertheless, simple surface
scattering models such as GOM and IEM allow us to explain fairly well
the variations of the actual radar scattering, as previously shown in
\cite{Paillou2014} for dunes only. These results confirm the
qualitative interpretation of the radar signature of linear dunes and
mega-yardangs presented in the previous sub-section.

\section{Application to Titan}
Comparative planetology is a powerful approach when studying remote
bodies, especially when only images of their surface are
available. This is particularly true for Titan, which shows a complex
and rich surface where geological and meteorological processes are
active, and where fieldwork is still a dream. It is even more
difficult to interpret radar images, which are less intuitive than
their optical counterparts, but we only have such (low resolution and
noisy) radar scenes to ``see'' the surface of Titan through its opaque
atmosphere. Studying analogue terrestrial landforms using a comparable
radar system (at least in terms of wavelength) can then give us some
useful indications and directions to help understand the Cassini RADAR
images. However, one has to remain cautious, since the actual origin,
age, morphology, and materials of the exotic structures observed on
the surface of Titan are not well known.  With this in mind, the
previous qualitative and quantitative study of the radar scattering of
terrestrial linear dunes and mega-yardangs can help us to investigate
similar linear structures on Titan. We focus here on some selected
sites on Titan imaged by the RADAR instrument, and show how the radar
images of terrestrial analogues can guide the interpretation.

\begin{figure*} \centering
  \includegraphics{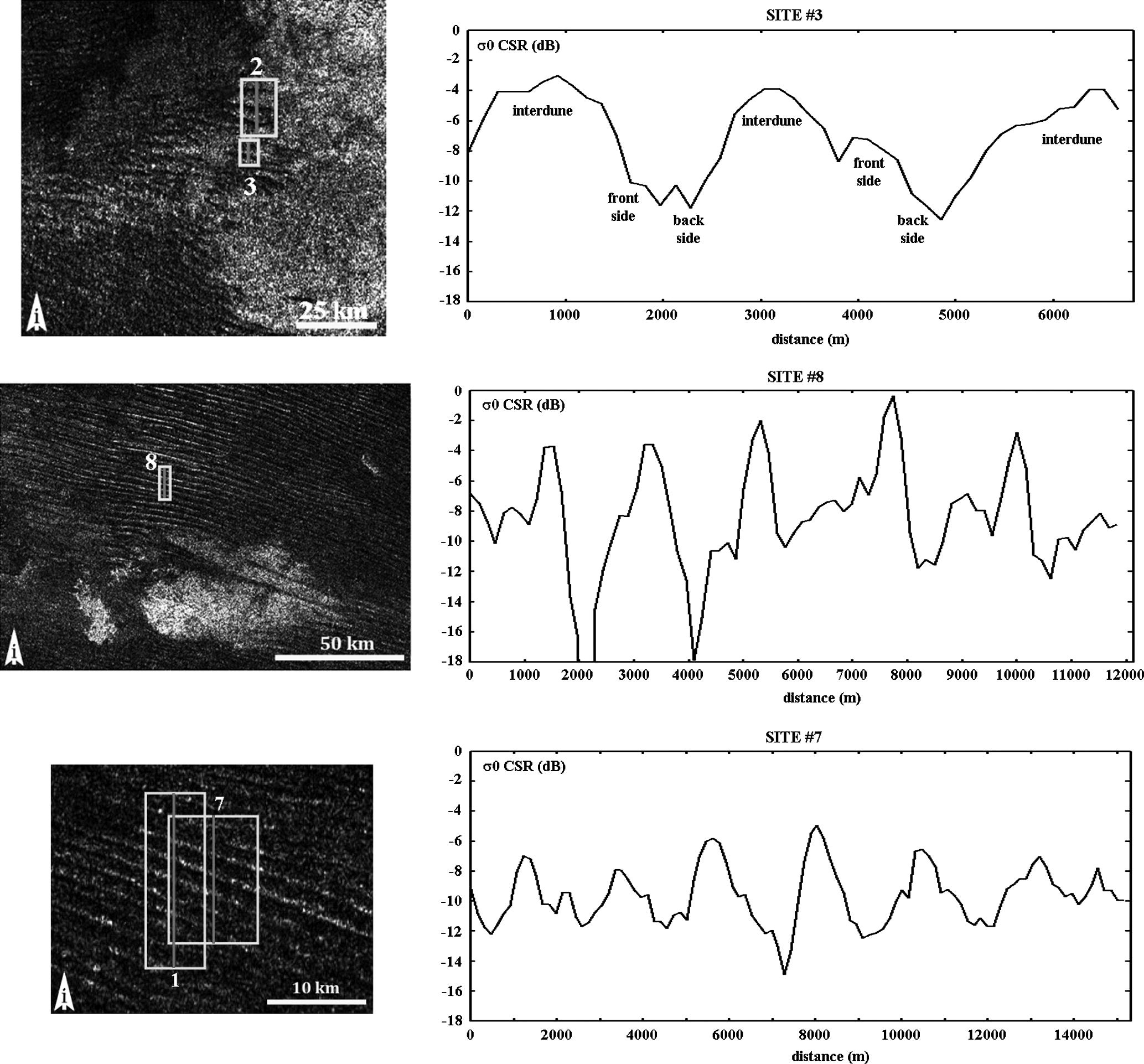}
  \caption{\label{fig:7}Extracts of Cassini RADAR flyby T8 showing
linear dunes of sites \#3, \#8, and \#7 with corresponding radar
scattering profiles. Site \#3 is located 6.5$^\circ$S/238.8$^\circ$W
with an incidence angle around 21$^\circ$, site \#8 is located
8.8$^\circ$S/259.4$^\circ$W with an incidence angle around 26$^\circ$,
and site \#7 is located 8.2$^\circ$S/248.9$^\circ$W with an incidence
angle around 26$^\circ$. North is down, the ``i'' arrow indicates
direction of the radar illumination.}
\end{figure*}

Regarding linear dunes, we considered the Cassini RADAR acquisition
performed during the T8 flyby in October 2005. This covered the more
than 2500 km-long Belet Sand Sea, a region dominated by linear dunes,
and stretching from 180$^\circ$W to 300$^\circ$W and from 20$^\circ$N
to 20$^\circ$S \citep{Lunine2008}.  Illumination direction is roughly
perpendicular to the orientation of the main axis of the dunes, so
that the acquisition geometry is comparable to that of TerraSAR-X
images presented in the previous section. Incidence angle during the
T8 flyby varied from 17$^\circ$ to 24$^\circ$, also comparable to the
range of incidence angle of our TerraSAR-X data.  The RADAR instrument
resolution is sufficiently coarse that it can be difficult to
determine if the linear dunes in the Belet Sand Sea have sandy
interdunes like those in Namibia \citep{Neish2010}, or are separated
by exposed bedrock like the Egyptian dunes \citep{Paillou2014}, or if
the two kinds can be found in different locations across Belet.  We
studied ten locations in the T8 flyby, where dune structures are
clearly imaged, and performed some averaging in the direction parallel
to the main orientation of the dunes in order to extract significant
radar scattering profiles. We present three interesting and typical
radar signatures in Fig.~\ref{fig:7}.  Site \#3 is located in a region
where radar-dark dunes end on what seems to be a radar-bright plateau,
and where the extracted radar scattering profile looks similar to
those observed for Egyptian dunes: a first radar-bright level (-4 dB)
corresponds to the rough interdune (the exposed surface of the
plateau), a second intermediate level (-7 to -8 dB) might correspond
to front sides of the dunes illuminated by the radar, and a third
lower level (-12 to -13 dB) could be associated to sides of the dunes
opposite to the radar illumination direction. Site \#8 is located in a
region that was previously studied by \cite{Neish2010} who applied
radarclinometry techniques to estimate dune heights. It shows the same
three ``Egyptian-like'' scattering levels: a radar-bright level (-1 to
-4 dB) that could again be associated with bare interdunes, an
intermediate level (-7 to -8 dB) that could correspond to front sides
of the dunes, and a lower scattering level ($<$-12 dB) that could be
associated with back sides of the dunes (and remains higher than the
Cassini RADAR noise level, around -18 dB). However, it is also
possible that dunes of site \#8 are of the Namibian type, and that the
radar-bright peaks are the sides of the dunes facing the direction of
the radar illumination, the intermediate level being the result of
interdune scattering, and the lower level being the back side of the
dunes. Interpreting the radar-bright features as interdunes
(Egyptian-type) or as sides of dunes facing the radar (Namibian-type)
will have a strong impact on the dunes’ shape estimation using
radarclinometry \citep{Neish2010}.  Site \#7 exhibits a more clear
Namibian-like bimodal radar signature. It shows alternation of
high-scattering return (-6 to -7 dB) and medium-scattering return
(around -12 dB): the signature of site \#7 looks very similar to that
of Namibian dunes, the radar-bright features being sides of the dunes
facing the radar, and the medium-scattering level being related to
sand-covered interdunes, mixed with the low return of back sides of
the dunes. One can also notice that the radar-scattering levels for
Titan’s dunes are generally higher than for terrestrial ones, possibly
due to some additional volume scattering in a heterogeneous or porous
material, as proposed by \cite{Paillou2014}.  While the low resolution
(at best 300 m) combined with the high noise level (around -18 dB) of
the RADAR instrument does not allow us to extract unambiguous radar
signatures, both scattering profiles of Egyptian and Namibian-type
dunes seem to be present in the Belet Sand Sea. Further studies of
Titan’s dune fields should then take into account such possible
different interpretations of the dark and bright linear features, and
evaluate their impact on the understanding of the morphology and
history of the structures.

\begin{figure*} \centering
  \includegraphics{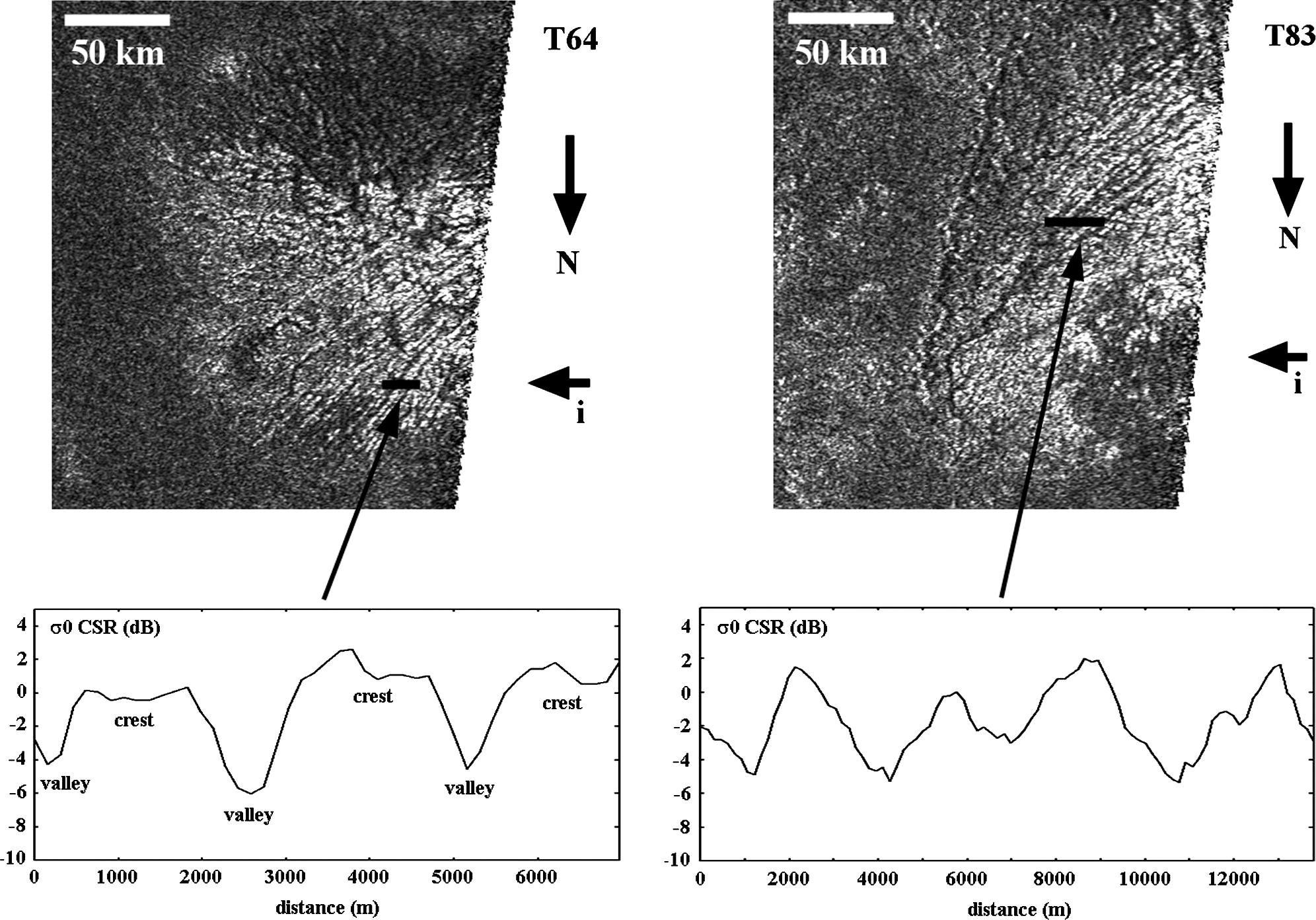}
  \caption{\label{fig:8} Extracts of T64 (left) and T83 (right)
Cassini RADAR acquisitions showing very bright linear structures,
interpreted as possible mega-yardangs, together with two studied radar
profiles. Resolution is 175 m/pixel, each image covers about 100 x 120
km, T64 yardang is located 41.3$^\circ$N/210.2$^\circ$W and T83
yardang is located 39.8$^\circ$N/196.6$^\circ$W, radar incidence angle
for both images is around 13$^\circ$.}
\end{figure*}

Concerning possible mega-yardangs on Titan, previous work already
proposed some candidate structures \citep{Paillou2013,Radebaugh2015}.
We considered two interesting landforms, observed at mid-latitudes
during the Cassini RADAR flybys T64 and T83. The T64 flyby was
performed in December 2009 (Cassini extended mission) and our region
of interest is centred on 41$^\circ$N/210$^\circ$W with an incidence
angle around 14$^\circ$. The T83 flyby was acquired in May 2012
(Cassini second extended mission) and our region of interest is
centred on 40$^\circ$N/197$^\circ$W with an incidence angle around
12$^\circ$. Fig.~\ref{fig:8} shows two very bright structures in T64
and T83, presenting alternating dark and bright linear features, which
are interpreted as possible mega-yardangs.  The two landforms are of
comparable size, and cover an area of about 60 x 60 km. Four selected
sites were studied in the T64 structure and three in the T83 one,
crossing bright-dark alternating linear features (see Fig.~\ref{fig:8}
presenting radar profiles across two representative sites). Again, we
averaged the radar scattering along the main direction of the linear
features, in order to derive significant radar scattering
profiles. While both T64 and T83 candidate mega-yargdangs look
similar, a closer look shows that the T64 structure is more like the
old mega-yardangs in Chad, while the T83 structure is more similar to
the Iranian mega-yardangs. Fig.~\ref{fig:8} presents the radar
scattering profile associated with site \#1 in T64 flyby: as for
Chadian mega-yardangs, two main scattering levels are observed, a very
bright level ($>$0 dB) corresponding to possible wide yardang crests,
and a less-bright level (around -5 dB) corresponding to possible
erosion valleys. Fig.~\ref{fig:8} also shows the radar profile
associated with site \#1 in T83 flyby: the same two radar scattering
levels as in T64 are observed there, but with a higher spatial
frequency alternation of narrower bright and less-bright features,
more similar to the signature of young yardangs observed in the Lut
Desert in Iran. We did not conduct any morphological measurements on
these potential extra-terrestrial yardangs to compare them to the
terrestrial ones (whose morphology is actually poorly described
\citep{Goudie2007}, but we rather focused here on their radiometric
signature. As for linear dunes, one can notice that the structures
observed on Titan are much brighter than their terrestrial analogues:
the brightest crests of Iranian and Chadian yardangs at around -5 dB
correspond to the low scattering level of the ``valleys'' of Titan’s
yardangs. Again, additional scattering processes, such as volume
scattering, should be considered on Titan. Nevertheless, even if the
global radar scattering level on Titan’s surface is higher, the
average scattering level difference between linear dunes and
mega-yardangs remains of the order of 10 dB, comparable to the
difference observed for their terrestrial analogues.  Finally, it is
interesting to note that the average radar scattering level of the T64
and T83 structures is around -2 dB (for an incidence angle around
13$^\circ$), a value comparable to the average scattering level of
north polar empty lakes basins \citep{Hayes2008,Michaelides2015}.  If
the radar-bright structures in T64 and T83 are mega-yardangs, they may
be the erosion result of unimodal wind direction over soft deposits
that could be ancient lake beds and/or evaporites
\citep{Barnes2011,Cordier2013}.  If confirmed, this would indicate the
possible existence of ancient lake basins at lower latitudes (around
40$^\circ$N) than polar regions in the past history of Titan.

\begin{figure} \centering
  \includegraphics{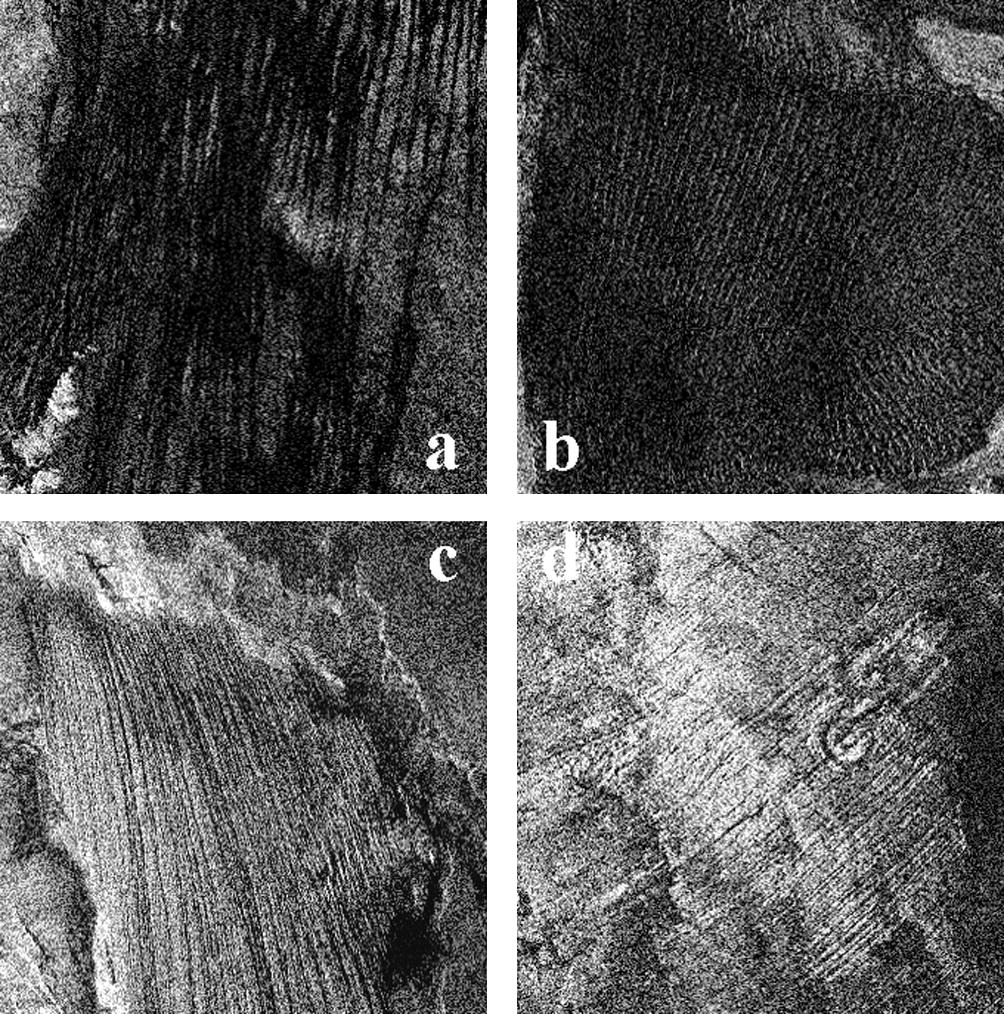}
  \caption{\label{fig:9} The four TerraSAR-X scenes of
Fig.~\ref{fig:1} with a resolution degraded to 300 m and a noise level
increased to -18 dB. Linear dunes in Egypt (a) and Namibia (b) look
quite similar to the structures observed in the Belet Sand Sea, and
mega-yardangs in Iran (c) and Chad (d) are recalling the radar-bright
structures observed in T64 and T83.}
\end{figure}

While comparing terrestrial geological landforms to ones observed on
Titan’s surface is not an easy exercise, the choice of relevant
structures on Earth and the use of a high-frequency orbital radar
sensor can nevertheless help to better interpret the noisy and low
resolution images provided by the Cassini RADAR instrument. In order
to illustrate, in a qualitative way, the similarity between the radar
signature of linear structures on Earth and Titan, we degraded the
four TerraSAR-X scenes that we previously used to a resolution and a
noise level similar to the RADAR instrument. Fig.~\ref{fig:9} presents
the resulting degraded images of linear dunes and mega-yardangs on
Earth, which are strongly reminiscent of the landforms seen on Titan’s
surface.
      
\section{Conclusion}
High resolution X-band radar images of terrestrial linear dunes and
mega-yardangs can help to better interpret the radar signature of
similar structures that are likely to exist on the surface of Titan.
Using a simple surface scattering model combined with topography data,
we accurately reproduced the radar backscattering profiles of linear
dunes (Egypt and Namibia) and mega-yardangs (Iran and Chad), thus
establishing a relationship between the radar radiometry and the
morphology and nature of imaged structures.

Applying this new understanding of the radar scattering process to the
interpretation of Cassini RADAR data, we have shown that both Egyptian
and Namibian-type dunes are present in the T8 flyby acquisition of the
Belet Sand Sea, i.e. interdune regions there can be either
sand-covered or not. This result should be taken into account for
future studies of the equatorial dune fields present on Titan.

We have also shown that two radar-bright structures, observed during
the T64 and T83 flyby acquisitions, are very likely to be the first
mega-yardangs observed on Titan. Such erosional structures could be
the remnants of past lake basins at mid-latitude, formed when Titan’s
climate was different, and need for further studies to understand
their origin.

As previously noticed by other authors, both dunes and yardangs on
Titan present a much stronger radar return than their terrestrial
counterparts, indicating that some additional scattering processes,
such as volume scattering in an heterogeneous or porous material,
occur on Titan.

\section*{Acknowledgments}

The authors would like to acknowledge the French space agency CNES for
providing financial support to Ph. Paillou and B. Seignovert for this
study, and the German space agency DLR for providing TerraSAR-X scenes
(proposal GEO1970). They also thank Tom Farr and Ralph Lorenz for
their advices and careful review of this paper.

\bibliography{dunes-yardangs-paillou}

\end{document}